\documentclass{PoS}
\usepackage{amsmath,amssymb,amsfonts,graphicx,slashed,booktabs}
\allowdisplaybreaks[1]

\title{Constraints on the chiral unitary {$\bar KN$} amplitude from 
{$\pi\Sigma K^+$} photoproduction data}

\ShortTitle{Constraints on the chiral unitary {$\bar KN$} amplitude from 
{$\pi\Sigma K^+$} photoproduction data}

\author{\speaker{Maxim Mai}
         \thanks{{I thank Ulf-G. Mei{\ss}ner for collaboration on the work reported here. This work is supported in part by DFG and NSFC (SFB/TR~110).}}\\
        Helmholtz--Institut f\"ur Strahlen- und Kernphysik (Theorie) and 
        Bethe Center for Theoretical Physics, Universit\"at Bonn, D-53115 Bonn, Germany\\
        E-mail: \email{mai@hiskp.uni-bonn.de}}

\abstract{A chiral unitary approach for antikaon-nucleon scattering in on-shell factorization is studied. We find multiple sets of parameters for which the model describes all existing hadronic data similarly well. We confirm the two-pole structure of the $\Lambda (1405)$. The narrow $\Lambda(1405)$ pole appears at comparable positions in the complex energy plane, whereas the location of the broad pole suffers  from a large  uncertainty. In the second step, we use a simple model for  photoproduction of $K^+\pi\Sigma$ off the proton and confront it with the experimental data from the CLAS collaboration. It is found that only a few of the hadronic solutions  allow for a consistent description of the CLAS data within the assumed reaction mechanism.}

\FullConference{The 8th  International Workshop on Chiral Dynamics \\
                 29 June 2015 - 03 July 2015\\
                 Pisa, Italy}

\begin{document}

\section{Introduction}

The strangeness $S=-1$ resonance $\Lambda(1405)$ is believed to be dynamically generated through coupled-channel effects in the antikaon-nucleon interaction. A further intricate feature is its two-pole structure. Within chiral unitary approaches, which are considered to be the best tool to address the chiral SU(3) dynamics in such type of system, the investigation of the two-pole structure was initiated in Ref.~\cite{Oller:2000fj} and thoroughly analyzed in many publications, for a review see Ref.~\cite{Hyodo:2011ur}. However, the scattering data alone do not allow to pin down the poles with good precision, as it is known since long, see e.g. Ref.~\cite{Borasoy:2006sr}.

Recently, very sophisticated measurements of the reaction $\gamma p\to K^+\Sigma \pi$ were performed by the CLAS collaboration at JLAB, see Ref.~\cite{Moriya:2013eb}. There, the invariant mass distribution of all three $\pi\Sigma$ channels was determined in a broad energy range and with high resolution. First theoretical analyses of this data have already been performed  on the basis of a chiral unitary approach in Refs.~\cite{Roca:2013av,Nakamura:2013boa}. In this work, we take up the challenge to combine our next-to-leading order approach of antikaon-nucleon scattering \cite{Mai:2012dt} in an on-shell approximation with the CLAS data. 

First, we construct a family of solutions that lead to a good description of the scattering and the SIDDHARTA data. This reconfirms the two-pole structure of the $\Lambda(1405)$. As before, we find that the location of the second pole in the complex energy plane is not well determined from these data alone. Then, we address the issue how this ambiguity can be constrained from the CLAS data. Similar to Ref.~\cite{Roca:2013av}, we use a simple semi-phenomenological model for the photoproduction process that combines the description of the hadronic scattering with a simple polynomial and energy-dependent ansatz for the photoproduction of $K^+$ and a meson-baryon pair of strangeness $S=-1$ off the proton. The corresponding energy- and channel-dependent constants are fit to the CLAS data. However, it appears that not all solutions, consistent with the scattering data, lead to a decent fit to the photoproduction data. Moreover, we find that the solutions, consistent with photoproduction and scattering data lead to 
similar positions of both poles of $\Lambda(1405)$.

\section{Antikaon-nucleon scattering}\label{sec:scat}

\subsection{Model}

The starting point of the present analysis is the meson-baryon scattering amplitude in the strangeness $S=-1$ sector. We assume a simplified version of the amplitude constructed and described in detail in the original publication \cite{Mai:2014xna} as well as in Refs.~\cite{Mai:2012wy,Bruns:2010sv}, to which we refer the reader for conceptual details. We start from the chiral Lagrangian of leading (LO) and next-to-leading (NLO) order. For the reasons given in Refs.~\cite{Mai:2014xna,Mai:2012wy,Bruns:2010sv}, the $s$- and $u$-channel one-baryon exchange diagrams are neglected, leaving us with the following chiral potential
\begin{align}\label{eqn:potential}
 V(\slashed{q}_2, \slashed{q}_1; p)=A_{WT}(\slashed{q_1}+\slashed{q_2}) +A_{14}(q_1\cdot q_2)+A_{57}[\slashed{q_1},\slashed{q_2}] +A_{M} +A_{811}\Big(\slashed{q_2}(q_1\cdot p)+\slashed{q_1}(q_2\cdot p)\Big)\,,
\end{align}
were the incoming and outgoing meson four-momenta are denoted by $q_1$ and $q_2$, whereas the overall four-momentum of the meson-baryon system is denoted by $p$. The $A_{WT}$, $A_{14}$, $A_{57}$, $A_{M}$ and $A_{811}$ are 10-dimensional  matrices which encode the coupling strengths between all 10 channels of the meson-baryon system for strangeness $S=-1$, i.e. $\{K^-p$, $\bar K^0 n$, $\pi^0\Lambda$, $\pi^0\Sigma^0$, $\pi^+\Sigma^-$, $\pi^-\Sigma^+$, $\eta\Lambda$, $\eta \Sigma^0$, $K^+\Xi^-$, $K^0\Xi^0\}$. These matrices  depend on the meson decay constants, the baryon mass in the chiral limit, the meson  masses as well as 14 low-energy constants (LECs) as specified the original publication~\cite{Mai:2014xna}.

Due to the appearance of the $\Lambda(1405)$ resonance just below the $\bar K N $ threshold and large momentum transfer, the strict chiral expansion is not applicable for the present system. Instead, the above potential is used as a driving term of the coupled-channel Bethe-Salpeter equation (BSE), for NLO approaches see e.g. Ref.~\cite{Mai:2012dt,Ikeda:2012au,Guo:2012vv,Borasoy:2006sr}. For the meson-baryon scattering amplitude $T(\slashed{q}_2, \slashed{q}_1; p)$ the integral equation to be solved reads
\begin{align}\label{eqn:BSE}
T(\slashed{q}_2, \slashed{q}_1; p)=V(\slashed{q}_2, \slashed{q}_1;p)   +i\int\frac{d^d l}{(2\pi)^d}V(\slashed{q}_2, \slashed{l}; p)   S(\slashed{p}-\slashed{l})\Delta(l)T(\slashed{l}, \slashed{q}_1; p)\,,
\end{align}
where $S$ and $\Delta$ represent the baryon (of mass $m$) and the meson (of mass $M$) propagator, respectively, and are given by $iS(\slashed{p}) = {i}/({\slashed{p}-m+i\epsilon})$
and $i\Delta(k) ={i}/({k^2-M^2+i\epsilon})$. Moreover, $T$, $V$, $S$ and $\Delta$ in the last expression are matrices in the channel space. The loop diagrams appearing above 
are treated using dimensional regularization and applying the usual $\overline{\rm MS}$ subtraction scheme in the spirit of our previous work \cite{Bruns:2010sv}. Note that the modified loop integrals are still scale-dependent. This scale $\mu$ reflects the influence of the higher-order terms not included in our potential. It is used as a fit parameter of our approach. To be precise, we have 6 such parameters in the isospin basis.

\begin{figure}[t]
\includegraphics[width=\linewidth]{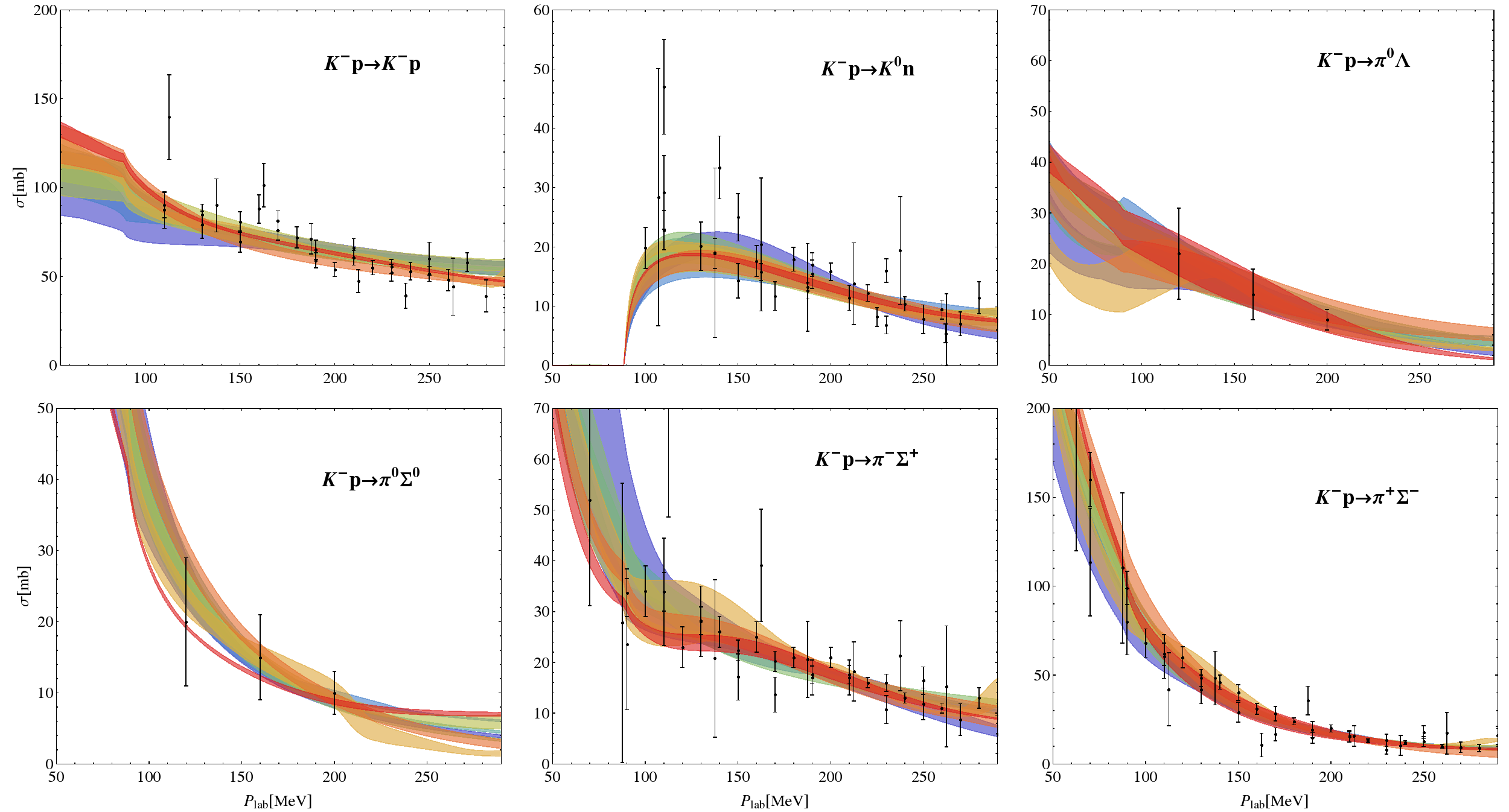}
\caption{Fit results compared to the experimental data from Refs.~\cite{Ciborowski:1982et,Humphrey:1962zz,Sakitt:1965kh,Watson:1963zz}. Different colors correspond to the eight best solutions, while the bands represent the $1\sigma$ uncertainty due to errors of the fit parameters. The color coding is specified in Fig.~\protect\ref{fig:poles1}.}\label{fig:cs}
\end{figure}

The above equation can be solved analytically, if the kernel contains contact terms only, see Ref.~\cite{Mai:2012wy} for the corresponding solution. Using this solution for the strangeness $S=-1$ system, we have shown in Ref.~\cite{Mai:2012dt} that once the full off-shell amplitude is constructed, one can easily reduce it to the on-shell solution, i.e. setting all tadpole integrals to zero. It appears that the double pole structure of the $\Lambda(1405)$ is preserved by this reduction and that the positions of the two poles are changing only by about $20$~MeV in imaginary part. On the other hand, the use of the on-shell approximation of the Eq.~\eqref{eqn:BSE} reduces the computational time roughly by a factor of 30. Therefore, since we wish to explore the parameter space in more detail, it seems to be safe and also quite meaningful to start from the solution of the BSE~\eqref{eqn:BSE} with the chiral potential \eqref{eqn:potential} on the mass-shell. Once the parameter space is explored well enough we can 
slowly turn on the tadpole integrals obtaining the full off-shell solution. Such a solution will become a part of a more sophisticated two-meson photoproduction amplitude in a future publication.

\subsection{Fit procedure and results}
\renewcommand{\baselinestretch}{1.25}
\begin{table}[t]
\begin{center}
\begin{tabular}{ccccccccc}
\toprule
Fit \# & 1&2&3&4&5&6&7&8\\
\midrule
$\chi_{\rm d.o.f.}^2$ (hadronic data) &1.35 &1.14&0.99 &0.96 &1.06 &1.02&1.15 &0.90\\
\midrule
$\chi_{\rm p.p.}^2$   (CLAS data)~~ &3.18&1.94&2.56&1.77&1.90&6.11&2.93&3.14\\
\bottomrule
\end{tabular} 
\caption{Quality of the various fits in the description of the hadronic and the photoproduction data from CLAS. For the definition of $\chi_{\rm p.p.}^2$, see the text.} \label{tab:photo}
\end{center}
\end{table}

The free parameters of the present model, the low-energy constants as well as the regularization scales $\mu$ are adjusted to reproduce all known experimental data in the meson-baryon sector of strangeness $S=-1$. The main bulk of this data consists of the cross sections for the processes $K^-p\to MB$, where $MB\in \{K^-p, \bar K^0n, \pi^0\Lambda, \pi^+\Sigma^-, \pi^0\Sigma^0, \pi^-\Sigma^+\}$, and laboratory momentum $P_{\rm lab}<300$ MeV, from Refs.~\cite{Ciborowski:1982et,Humphrey:1962zz,Sakitt:1965kh,Watson:1963zz}. Electromagnetic effects are not included in the analysis and assumed to be negligible at the measured values of $P_{\rm lab}$. Additionally, at the antikaon-nucleon threshold, we consider the decay ratios from Refs.~\cite{Tovee:1971ga,Nowak:1978au} as well as the energy shift and width of kaonic hydrogen in the 1s state from the SIDDHARTA experiment at DA$\Phi$NE \cite{Bazzi:2011zj} related to the $K^-p$ scattering length via the modified Deser-type formula \cite{Meissner:2004jr}. Due to 
the precision of the experiment, the latter two values have already become the most important input in this sector. In principle, both $\bar KN$ scattering lengths can be determined directly, performing a complementary measurement on the kaonic deuterium, see Refs.~\cite{LNF,JPARK} for the proposed experiments. The strong energy shift and width of the latter can again be related to the antikaon-deuteron scattering length, using the the modified Deser-type formula \cite{Meissner:2004jr} and finally to the antikaon-nucleon scattering lengths as described in Ref.~\cite{Mai:2014uma}.

The fit to the above data was performed minimizing $\chi_{\rm d.o.f.}^2$ using several thousands randomly distributed sets of starting values of the free parameters. The latter were assumed to be of natural size, while the unphysical solutions, e.g. poles on the first Riemann sheet for ${\rm Im}(W)<200$ MeV ($W:=\sqrt{p^2}$), were sorted out. For more details on the fitting procedure and results, we refer the reader to the original publication \cite{Mai:2014xna}. Eight best solutions were obtained by this, see second row of Tab.~\ref{tab:photo}, whereas the next best $\chi_{\rm d.o.f.}^2$ are at least one order of magnitude larger. The results of the fits compared to experimental data are presented in Fig.~\ref{fig:cs}, where every solution is represented by a distinct color.

\begin{figure}[t]
\begin{center}
\includegraphics[width=\linewidth]{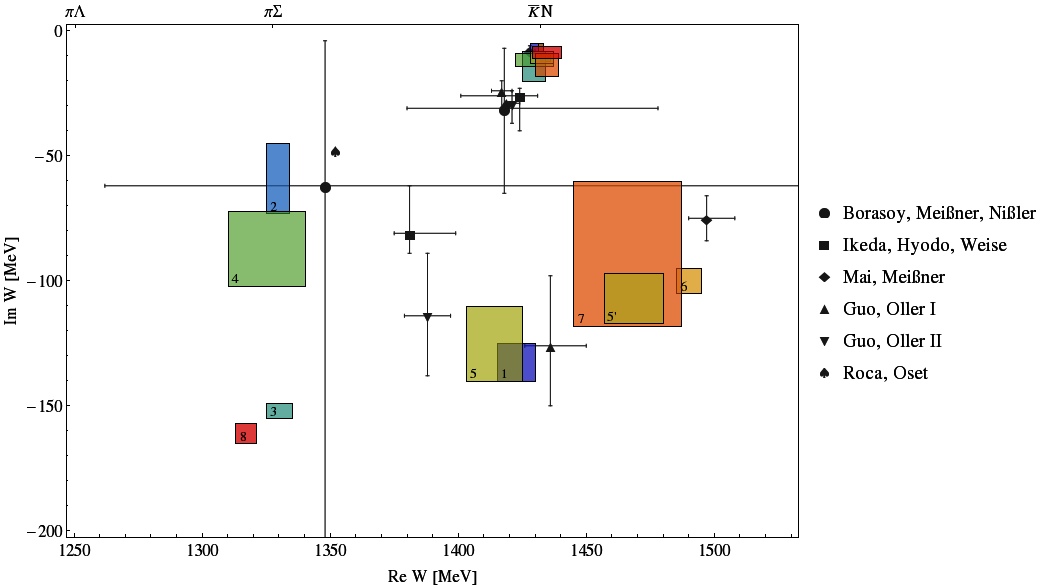}
\caption{Double pole structure of the $\Lambda(1405)$ in the complex energy plane for the eight solutions that describe the scattering and the SIDDHARTA data. For easier reading,  we have labeled the second pole of these solutions by the corresponding fit \#, where $5$ and $5'$ denote the second pole on the second Riemann sheet, connected to the real axis between the $\pi\Sigma-\bar K N$ and $\bar K N-\eta\Lambda$ thresholds, respectively. For comparison, various results from the literature are also shown, see Refs.~\protect\cite{Borasoy:2006sr,Guo:2012vv,Ikeda:2012au,Mai:2012dt,Roca:2013av}. }
\label{fig:poles1}                                                                                
\end{center}
\end{figure}
The data are described equally well by all eight solutions, showing, however, different functional behaviour of the cross sections as a function of $P_{\rm lab}$. When continued analytically to the complex $W$ plane, all eight solutions confirm the double pole structure of the $\Lambda(1405)$, see Fig.~\ref{fig:poles1}. There, the narrow pole lies on the Riemann sheet, connected to the real axis between the $\pi\Sigma-\bar KN$ thresholds for every solution. The second poles lie on the Riemann sheets, connected to the real axis between the following thresholds: $\pi\Sigma-\bar KN$ for solution 1, 2, 4, 5  and 8; $\pi\Lambda-\pi\Sigma$ for solution 3; $\bar K N-\eta\Lambda$ for solutions 6 and 7. Please note that the second pole of the solution 5 has a shadow pole (5' in Fig~\ref{fig:poles1}) on the Riemann sheet, connected to the real axis between $\bar K N-\eta\Lambda$ thresholds. The scattering amplitude is restricted around the $\bar K N$ threshold by the SIDDHARTA measurement quite strongly. Therefore, in 
the complex $W$ plane we observe a very stable behaviour of the amplitude at this energy, i.e. the position of the narrow pole agrees among all solutions within the $1\sigma$ parameter errors, see Fig.~\ref{fig:poles1}. This is in line with the findings of other groups~\cite{Ikeda:2012au,Borasoy:2006sr,Guo:2012vv}, i.e. one observes stability of the position of the narrow pole. Quantitatively, the first pole found in these models is located at somewhat lower energies and is slightly broader than those of our model. In view of the stability of the pole position, we trace this shift to the different treatment of the Born term contributions to the chiral potential utilized in Refs.~\cite{Ikeda:2012au,Borasoy:2006sr,Guo:2012vv}.

The position of the second pole is, on the other hand, less restricted. To be more precise, for the real part we find three clusters of these poles: around the $\pi\Sigma$ threshold, around the $\bar K N$ threshold as well as 
around $1470$~MeV. For several solutions there is some agreement in the positions of the second pole between the present analysis and the one of Ref.~\cite{Guo:2012vv} and of our previous work \cite{Mai:2012dt}. However, as the 
experimental data is described similarly well by all fit solutions, one  can not reject any of them. Thus, the distribution of poles represents the systematic uncertainty of the present approach.  It  appears to be quite large, but is still significantly smaller than the older analysis of Ref.~\cite{Borasoy:2006sr}. Recall that no restrictions were put on the  parameters of the model, except for naturalness.

\section{Photoproduction amplitude}
\begin{figure}[t]
\begin{center}
\includegraphics[width=0.99\linewidth]{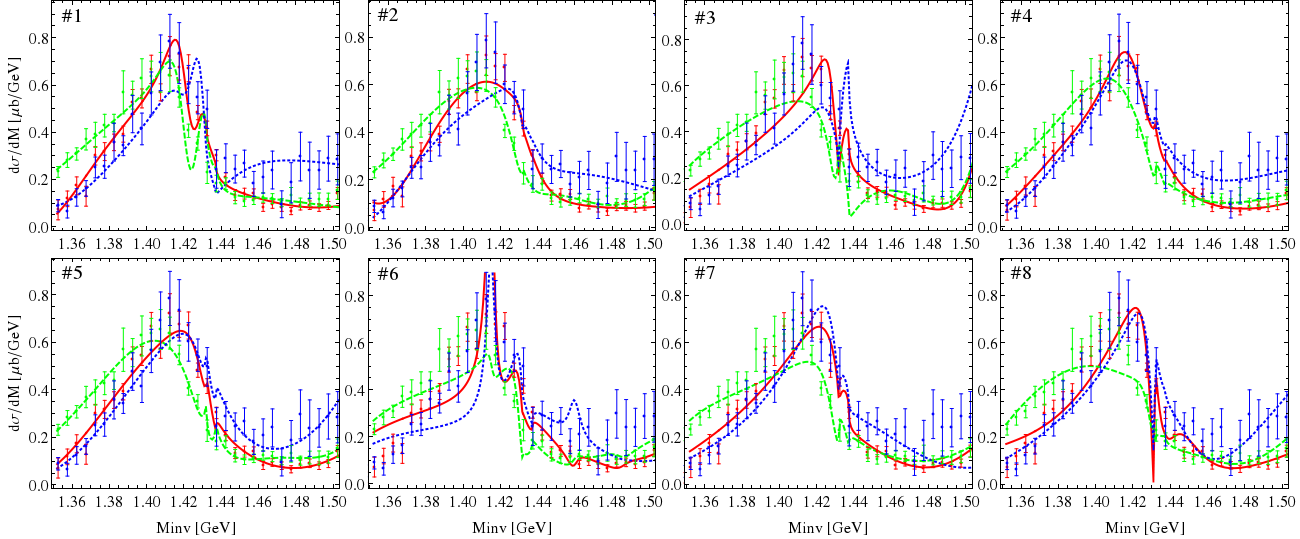}
\caption{Comparison of all solutions describing the $\pi \Sigma$ 
mass distribution at $\tilde W=2.5$ GeV in all three channels $\pi^
+\Sigma^-$ (green, dashed), $\pi^-\Sigma^+$ (full, red) and  $\pi^0\Sigma^0$ (blue, dotted).}
\label{pic:AUSSCHLUSS}
\end{center}
\end{figure}

In the last section we have demonstrated that the present model for the meson-baryon interaction possess at least eight different solutions, which all describe the hadronic data similarly well. In this section, we wish to see whether these solutions are compatible with the photoproduction data, if they are considered as a final-state interaction of the reaction  $\gamma p \to K^+\Sigma \pi$. For this purpose it is sufficient to consider the simple ansatz
\begin{equation}\label{eq:photo}
\mathcal M^j(\tilde W,M_{\rm inv}) 
= \sum_{i=1}^{10} C^i(\tilde W)\,G^i(M_{\rm inv})\,f_{0+}^{i,j}(M_{\rm inv})\,,
\end{equation}
where $\tilde W$ and $M_{\rm inv}$ denote the total energy of the system and the invariant mass of the $\pi\Sigma$ subsystem, respectively. For a specific meson-baryon channel $i$, the energy-dependent (and in general complex valued) constants  $C^i(\tilde W)$ describe the reaction mechanism of $\gamma p\to K^+M_iB_i$, where\-as the final-state interaction is captured by the standard H\"oh\-ler partial waves $f_{0+}$. For a specific meson-baryon channel $i$, the Greens function is denoted by $G^i(M_{\rm inv})$ and is given by the one-loop meson baryon function in dimensional regularization.

The regularization scales appearing in the Eq.~(\ref{eq:photo}) via the $G^i(M_{\rm inv})$ have already been fixed in the fit to the hadronic cross sections and the SIDDHARTA data.  Thus, the only new parameters of the photoproduction amplitude are the constants $C^i(\tilde W)$ which, however, are quite numerous (10 for each $\tilde W$). These parameters are adjusted to reproduce the invariant mass distribution $d\sigma/dM_{\rm inv}(M_{\rm inv})$ for the final $\pi^+\Sigma^-$, $\pi^0 \Sigma^0$ and $\pi^-\Sigma^+$ states and for all 9 measured total energy values $\tilde W=2.0,\,2.1,\,..,\,2.8$~GeV. The achieved  quality  of the photoproduction fits is listed in the third row of Tab.~\ref{tab:photo}, whereas the $\chi_{\rm d.o.f.}^2$ of the hadronic part are stated in the second row. Note that for the comparison of the photoproduction fits the quantity $\chi_{\rm d.o.f.}^2$ is not meaningful due to the large number of generic parameters $C_i(\tilde W)$. Therefore, we compare the total $\chi^2$ divided by the 
total number of data points for all three $\pi\Sigma$ final states, denoted by $\chi_{\rm p.p.}^2$. For the same reason it is not meaningful to perform a global fit, minimizing the total $\chi_{\rm d.o.f.}^2$. 

It turns out that even within such a simple and flexible photoproduction amplitude, only the solutions~\#2, \#4 and \#5 of the eight hadronic solutions allow for a decent description of the CLAS data. While the total $\chi^2$ per data point of these solutions is very close to each other, the next best solution has a 40\% larger total $\chi^2_{\rm p.p.}$ than the best one. The failure of the solutions \#1, \#3, \#6, \#7 and \#8 becomes quite evident in a one-to-one comparison of all eight solutions fitted to the CLAS data as presented in Fig.~\ref{pic:AUSSCHLUSS} for one particular cms energy chosen as a typical example. Moreover,  the hadronic amplitudes are determined up to 1$\sigma$ error bands. Therefore, it is a priori not clear, whether some of the hadronic solutions lying within these error bands may lead to a better fit of the CLAS data. We have checked this explicitly, considering a large number of hadronic solutions distributed randomly around the central ones. For every such solution a fit to the CLAS data was performed independently and no significantly better fit was found to those of the central solution. Therefore, we consider the above exclusion principle of the hadronic solutions as statistically stable. For further discussion on this aspect see Ref.~\cite{Mai:2014xna}.

The best solution is indeed \#4, which, incidentally, has also the lowest $\chi_{\rm d.o.f.}^2$ for the hadronic part. This solution also gives an excellent description of the $\Sigma \pi\pi$ mass distribution from Ref.~\cite{Hemingway:1984pz}, calculated using the method developed in Ref.~\cite{Oller:2000fj}. With respect to these data, solution \#2 is also satisfactory but \#5 is not. Therefore, the photoproduction data combined with the scattering and the SIDDHARTA data lead to a sizable reduction in the ambiguity of the second pole of the $\Lambda (1405)$. In fact, the second pole of the surviving solutions is close to the value found in Ref.~\cite{Roca:2013av}, see Fig.~\ref{fig:poles1}, and also close to the central value of the analysis based on scattering data only \cite{Borasoy:2006sr}. To be precise, the locations of the two poles in these surviving solutions are (all in MeV)

\begin{center}
\renewcommand{\baselinestretch}{1.5}
\begin{tabular}{ccc}
\toprule
solution & pole 1 & pole 2 \\
\midrule
\#2  & $1434^{+2}_{-2} - i \, 10^{+2}_{-1}$ & $1330^{+4~}_{-5~} - i \, 56^{+17}_{-11}$\\
\#4  & $1429^{+8}_{-7} - i \, 12^{+2}_{-3}$ & $1325^{+15}_{-15} - i \, 90^{+12}_{-18}$\\
\bottomrule
\end{tabular}\end{center}

We conclude that the inclusion of the CLAS data as experimental input can serve as a new important constraint on the antikaon-nucleon scattering amplitude. However, for  future studies a theoretically more robust model for the two-meson photoproduction amplitude is required. We propose that a generalization of the one-meson photoproduction model, presented in Ref.~\cite{Borasoy:2007ku,Mai:2012wy}, may be the next logical step for this endeavor.

\newpage


\end{document}